\documentclass[twocolumn,prl,aps,showpacs,superscriptaddress]{revtex4}
\usepackage{amsmath}
\usepackage{graphicx}
\usepackage[usenames]{color}

\usepackage[ps2pdf,colorlinks,bookmarks]{hyperref}
\definecolor{linkblue}{rgb}{0,0,0.8}
\definecolor{linkgreen}{rgb}{0,0.5,0}
\hypersetup{pdfpagemode=None, pdfstartview=FitH, linkcolor=linkblue, %
             citecolor=linkgreen, urlcolor=linkblue}

\def\beq{\begin{equation}}
\def\eeq{\end{equation}}
\def\bea{\setlength\arraycolsep{1.4pt}\begin{eqnarray}}
\def\eea{\end{eqnarray}}
\def\bit{\begin{itemize}}
\def\eit{\end{itemize}}

\def\eqs{Eqs.~}
\def\fig{Fig.~}

\def\ld{\left}
\def\rd{\right}

\def\fr{\frac}

\def\Lam{\Lambda}

\def\P{{\cal P}}
\def\R{{\cal R}}

\def\R{{\cal R}}
\def\PR{{\cal P_{\cal R}}}

\def\LCDM{$\Lambda$CDM}
\def\Planck{\textit{Planck}}

\def\LLTB{$\Lambda$LTB}

\begin{document}

\title{Nowhere to hide: closing in on cosmological homogeneity}

\author{James P. Zibin} \email{zibin@phas.ubc.ca}
\affiliation{Department of Physics \& Astronomy\\
University of British Columbia, Vancouver, BC, V6T 1Z1  Canada}

\author{Adam Moss} \email{adam.moss@nottingham.ac.uk}
\affiliation{Centre for  Astronomy \& Particle Theory \\ University of 
Nottingham, University Park, Nottingham, NG7 2RD, U.K.}

\date{\today}

\begin{abstract}

   Homogeneity is a crucial, but poorly tested, assumption in cosmology.  
We introduce a new approach which allows us to place limits on the 
presence of localized structures within essentially our entire observable 
volume, using cosmic microwave background secondary anisotropies.  
We find that structures cannot exceed roughly 20 times their expected 
amplitude over most of our observable volume.  
Similarly, we place tight constraints on {\em statistical} inhomogeneity 
within our volume, performing the first power spectrum reconstruction 
using secondary anisotropies alone.  We find that the standard model passes 
this important new consistency test.  Our approach probes homogeneity 
over vastly larger volumes and scales than previous studies based on surveys.

\end{abstract}

\maketitle


\section{Introduction}

   The standard $\Lam$ cold dark matter (\LCDM) 
cosmological model rests upon a few essential foundations.  First, it assumes 
that Einstein's general relativity (GR) is valid.  Second, it takes the 
matter content to be a mixture 
of radiation, baryons, and CDM, together with a cosmological constant.  
Finally, it adopts the cosmological principle, namely, that the Universe 
approaches homogeneity and isotropy on the largest scales, and also assumes 
that the primordial fluctuations are {\em statistically} homogeneous and 
isotropic.  The first two foundations have been intensively examined, via 
modified gravity models (see, e.g., \cite{cfps12} for a review) and via 
dark energy models (see \cite{cst06,llww11} for reviews) and non-standard 
neutrino and dark matter content.

   However, the assumption of homogeneity has received considerably 
less critical examination.  One reason for this is the familiar textbook 
argument (see, e.g., \cite{wald84}) that homogeneity is implied by isotropy 
of the spacetime [sometimes taken to be implied by isotropy of the cosmic 
microwave background (CMB) \cite{noteisotr}] together with the assumption of 
the Copernican principle (so that the Universe appears isotropic to {\em all} 
observers) \cite{noteinfl}.  But we can circumvent this argument in two 
ways.  First, if we allow for violations of the Copernican principle, then 
{\em radial} inhomogeneity centred on us will still be consistent with 
isotropy.  Such models received considerable interest in recent years as 
alternatives to accelerating models (see, e.g., \cite{clarkson12} for a 
review), although ultimately were unsuccessful when confronted with a 
variety of observational probes 
\cite{bnv10,mzs11,zs11,zibin11,bcf12,zgbrl12,planck13kSZ,rbmlb14}.  However, 
more subtle radial inhomogeneity may still bias the determination of 
cosmological parameters \cite{davisetal11,vb12,vmc14,vkm13,wojtaketal13}.  
Nevertheless, we will not consider anti-Copernican models in this paper.

   The second way to allow for inhomogeneity is to simply note that the 
CMB (and our spacetime) is {\em not,} of course, perfectly isotropic; 
rather, it contains fluctuations on many scales with amplitudes of order 
$10^{-5}$.  Therefore inhomogeneities (beyond those expected in \LCDM) 
may be present between us and our last scattering surface (LSS) if 
they produce sufficiently weak secondary CMB anisotropies \cite{noteassump}.  
The question then becomes: how large might such inhomogeneities be, 
without producing {\em obvious} signs in the CMB?  This is a central question 
that we address in this paper.  Clearly, the implications of any such 
violations of homogeneity could be profound.

   Consistency between observations and homogeneity has been examined before, 
but almost always using surveys to map large-scale structure (see, e.g., 
\cite{hoggetal05,ybps05,sypb09,hirata09,mbb12,scrimgeouretal12,hoyleetal12}).  
These studies have 
been consistent with homogeneity; claims of anomalously large structures 
\cite{sd11,clowesetal13} have subsequently been refuted 
\cite{parketal12,nadathur13,pm13}.  However, current galaxy surveys reach 
only to redshifts $z \simeq 1$, hence sampling only $\sim$$1\%$ of our 
observable comoving volume.  Limited sky coverage further reduces this 
volume.  Some studies \cite{mgs95,bfs96,khb97,sag97,planck13bgndtopo} have 
used CMB data to constrain particular inhomogeneous or anisotropic models, 
usually Bianchi models.  
Ref.~\cite{vmc14} used multiple probes 
to constrain inhomogeneity, but only considered radial structures, i.e., 
tested for violations of the Copernican principle.

   The goal of this paper is to extend the range of scales and 
redshifts sampled by surveys to test homogeneity over most of our 
observable volume, using the secondary CMB probes of the integrated 
Sachs-Wolfe (ISW) effect, the kinetic Sunyaev-Zeldovich (kSZ) effect, 
the Rees-Sciama (RS) effect, and CMB lensing, as well as galaxy 
lensing~\cite{notehomog}.  To do this we must choose parameterizations for 
the inhomogeneities.  We consider spherical structures (motivated by various 
models for inflationary artifacts \cite{fik10,asw11,amy13}) and a 
statistically inhomogeneous power spectrum (motivated by hints of 
large-scale CMB power asymmetries \cite{ehbgl04,planck13isostat}).  
Constraints in the latter case amount to a CMB-secondary-based primordial 
power spectrum reconstruction.  In both cases we consider an otherwise 
standard \LCDM\ universe with \Planck\ best-fit parameters 
\cite{planck13params}.  Note that our 
goal is {\em not} to perform a targeted search for structures, as has 
been done for particular inflationary relics 
\cite{fjmmp12,amy13}.  Instead, we wish to place upper limits on the 
amplitudes of structures that would be {\em undetectable} against the CMB 
fluctuations, and 
hence place limits on allowed departures from homogeneity.


\section{Cosmological probes}

   Our primary goal in this paper is to 
constrain homogeneity over the largest volumes accesible to us.  To do 
this, we must use probes which reach as far as possible.  Ideally, we 
would also prefer methods sensitive to total matter (i.e., metric) 
fluctuations, to avoid bias uncertainties.  The ISW, kSZ, RS, and CMB and 
galaxy lensing probes satisfy these criteria to various degrees.

   To characterize the usefulness of these various observations, we 
will express their power spectra as line-of-sight 
integrals in the Limber approximation \cite{limber53}.  
First, the ISW spectrum can be written
\beq
\ell^2C_\ell^{\rm ISW} \simeq \fr{72\pi^2}{25\ell}
                              \int_0^{r_{\rm LS}}dr\,r\,{g'}^2(r)
                       \PR\ld(\fr{\ell}{r}\rd)T^2\ld(\fr{\ell}{r}\rd).
\label{ClISWlimb}
\eeq
Here $r_{\rm LS}$ is the comoving radius to the LSS, $g$ the $\Lam$ growth 
supression factor, $\PR(k)$ the (dimensionless) primordial comoving 
curvature spectrum, $T(k)$ the (linear) transfer function 
accounting for the suppression during radiation domination, $k$ the comoving 
wavenumber, and a prime indicates a conformal time derivative.

   Similarly, we can write the CMB lensing potential power spectrum in the 
Limber approximation as \cite{lc06}
\beq
\ell^4C_\ell^{\phi} \simeq \fr{72\pi^2\ell}{25}
   \!\!\int_0^{r_{\rm LS}}\!\!\!\!\!\!\!\!\!dr r\!
   \ld(\fr{r_{\rm LS} - r}{r_{\rm LS}r}\rd)^2\!\!\!g^2(r)
   \PR\!\!\ld(\fr{\ell}{r}\rd)\!T^2\!\!\ld(\fr{\ell}{r}\rd).
\label{Cllenslimb}
\eeq
Finally, the galaxy lensing convergence power spectrum for sources at a 
single distance $r_{\rm s}$ becomes (see, e.g., \cite{huterer02})
\beq
\ell^2C_\ell^{\kappa} \simeq \fr{18\pi^2\ell^3}{25}
   \!\!\int_0^{r_{\rm s}}\!\!\!\!\!\!dr r\!
   \ld(\fr{r_{\rm s} - r}{r_{\rm s}r}\rd)^2\!\!\!g^2(r)
   \PR\!\!\ld(\fr{\ell}{r}\rd)\!T^2\!\!\ld(\fr{\ell}{r}\rd).
\label{Clgallenslimb}
\eeq

\begin{figure}
\includegraphics[width=\columnwidth]{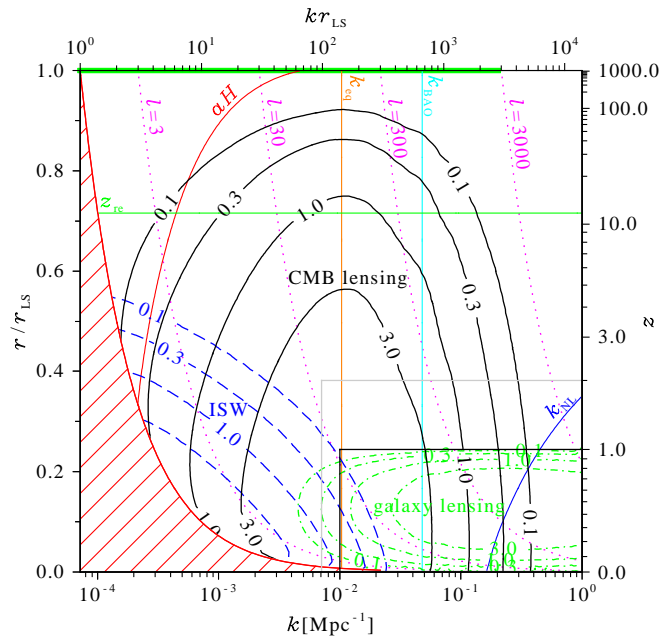}
\caption{Limber approximation kernels for the CMB lensing potential 
(solid black contours), ISW effect (dashed blue contours), and galaxy 
lensing (dot-dashed green contours).  Normalization is arbitrary.  The 
dotted magenta curves indicate the corresponding Limber multipole scale, 
$\ell = kr$.  Scales 
labelled $aH$, $k_{\rm eq}$, $k_{\rm BAO}$, and $k_{\rm NL}$ are, 
respectively, the comoving Hubble scale, the equality scale, the first 
baryon acoustic oscillation peak, and the nonlinearity scale, beyond which 
information is harder to extract.  Geometry prevents measurements in the 
hatched region (delimited roughly by $\ell = 1$).  The black box roughly 
indicates the range accessed by the WiggleZ survey 
\cite{scrimgeouretal12,pooleetal13}, the grey box by the proposed 
Euclid survey \cite{cimattietal09}, and the heavy green line by primary CMB.
\label{krplot}}
\end{figure}

   To get an idea of the sensitivity of these probes to the scales 
and redshifts of fluctuations, we plot in \fig\ref{krplot} the kernels 
of \eqs(\ref{ClISWlimb})--(\ref{Clgallenslimb}) in the $k$-$r$ plane, 
together with the regions sampled by galaxy surveys and measurements of 
the primary CMB \cite{noteell}.  For galaxy lensing we choose 
$r_{\rm s} = r(z = 1)$.  It is clear from this plot that current galaxy 
surveys sample only a small fraction of the comoving distance to last 
scattering, and also are insensitive to the largest two decades of length 
scales that are in principle observable.  Similarly, while the primary 
CMB samples a large range of scales, it is sensitive to distances very 
close to last scattering ($r_{\rm LS} - r \lesssim r_{\rm LS}/1000$).  
Therefore, to the extent that galaxy surveys and primary CMB have been 
utilized to test 
homogeneity, there is much more room for departures from homogeneity to 
``hide''.  In particular, modifications to the matter power spectrum or 
localized structures on scales $10^{-4}\,{\rm Mpc}^{-1} < k < 10^{-2}\,
{\rm Mpc^{-1}}$, or inhomogeneities at $1 \lesssim z \lesssim 1000$ are 
possible while maintaining consistency with surveys and primary CMB.  Any 
such power spectrum modifications must be localized away from the LSS 
for consistency with the primary CMB, so they would entail a 
breaking of statistical homogeneity.  Figure~\ref{krplot} makes it apparent 
that ISW and especially CMB lensing are the most promising probes of 
the region not accessible to galaxy surveys.

\section{Localized structures}

   Next we consider the matter of constraining 
localized structures that may be ``lurking'' outside the reach of surveys but 
inside our LSS.  We will see that even some {\em nonlinear} structures will 
be allowed, so we will treat the structure with exact GR.  We use 
the spherically symmetric $\Lam$-Lema\^itre-Tolman-Bondi (\LLTB) spacetime 
\cite{omer65}, sourced by dust and $\Lam$, to represent the standard \LCDM\ 
background with superposed spherical structure \cite{noteradn}.  We choose a 
compensated underdensity with \LLTB\ curvature function profile (in the 
notation of \cite{zibin11})
\beq
K(r) = 
  K_0\ld[2\ld(\fr{r}{R}\rd)^5 - 3\ld(\fr{r}{R}\rd)^4 + \ld(\fr{r}{R}\rd)^2\rd],
\label{profile}
\eeq
with amplitude $K_0$ and comoving radius $R$.  The ISW, RS, and SW temperature 
anisotropies due to the structure are calculated by evolving null 
geodesics from the observer to the LSS, as described in 
\cite{zm11,zibin11}.  The deflection angles, 
$\boldsymbol{\alpha}$, are translated into lensing potential, $\phi$, via 
$\boldsymbol{\alpha} = \boldsymbol{\nabla}\phi$.  The \LLTB\ solution is 
calculated by numerically evolving Einstein's equations using 
independent formulations as checks, including that descibed in 
\cite{zibin08}.   We also monitor the constraints and 
compare with LTB and linear theory in the appropriate regimes.  The 
kSZ anisotropies are calculated using linear theory (with reionization width 
$\Delta z = 0.5$), which we confirm to be in good agreement with the 
much more time-consuming \LLTB\ kSZ calculation where the kSZ signal 
dominates.  We calculate only the kSZ effect due to the structure itself, 
not to smaller superimposed structures.

   We vary the observer's position, $r$, and the structure's radius.  
For each pair $(R,r)$, we iterate to find the amplitude $K_0$ that results 
in unity $S/N$ of the structure against the CMB fluctuations, where
\beq
\ld(\fr{S}{N}\rd)^2 = \sum_{\ell m} \fr{|a_{\ell m}^{T}|^2}{C_\ell^{T}}
        + \sum_{\ell m} \fr{|a_{\ell m}^{\phi}|^2}{C_\ell^{\phi} + N_\ell}.
\eeq
Here $a_{\ell m}^{T}$ and $a_{\ell m}^{\phi}$ are the temperature and lensing 
potential multipoles due to the structure, $C_\ell^{T}$ and 
$C_\ell^{\phi}$ are the standard \LCDM\ power spectra, and $N_\ell$ is 
the \Planck\ lensing reconstruction noise~\cite{notePLA}.  The temperature 
contribution is summed over $2 \le \ell \le 2000$, while for lensing we 
restrict the sum to $40 \le \ell \le 400$, corresponding to the range reliably 
measured by \Planck\ \cite{planck13lensing,notelens}.

\begin{figure}
\includegraphics[width=\columnwidth]{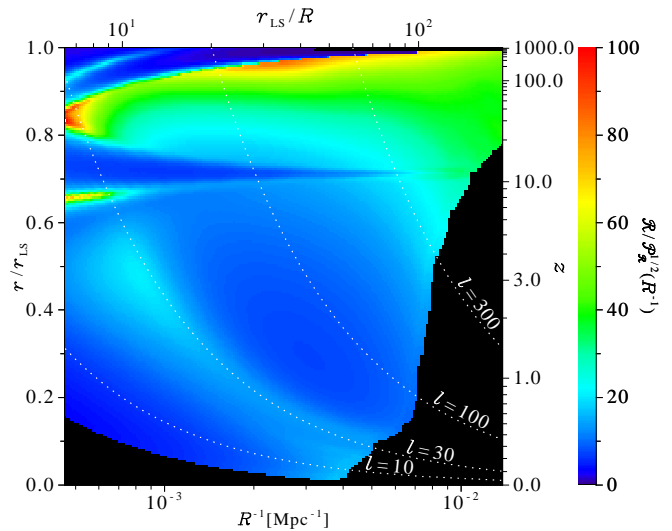}
\caption{Largest undetectable amplitude for localized structures relative 
to the expected \LCDM\ amplitude.  The structures' approximate peak multipole 
scales are also indicated.  Constraints are tightest (blue regions) near 
$r = r_{\rm LS}$ 
(due to the SW effect), near reionization (kSZ), in the lower left corner 
(ISW), and in the centre (CMB lensing plus subdominant RS).  In the black 
area at the bottom the observer would be inside the structure, and in the 
black area at right shell crossings occur before $S/N = 1$ is reached.
\label{psilimplot}}
\end{figure}

   The results are presented in \fig\ref{psilimplot}, where the colour 
values indicate the structure amplitude required to give a total $S/N$ of 
unity.  The plotted value is the ratio of the perturbation amplitude, $\R$, 
to the amplitude {\em expected} in \LCDM, $\PR^{1/2}(R^{-1})$ 
\cite{noteampl}.  Thus a colour value of, e.g., 
$10$ at some $(R,r)$ means that a structure at that distance and radius must 
have an amplitude less than $10$ times the expected \LCDM\ value in order 
to be undetectable against the CMB.  Comparing Figs.~\ref{psilimplot} and 
\ref{krplot}, we can see that we have tighter constraints in the regions 
corresponding to the ISW and CMB lensing, and, as expected, very tight 
constraints close to the LSS, where the SW effect dominates.  (Note 
that we can roughly relate $k \simeq 2\pi/R$.)  A wedge near reionization 
shows that the kSZ is a strong constraint there.  The majority of the 
$(R,r)$ plane is constrained most tightly by CMB lensing.  Some of the 
parameter space 
is untestable with the \LLTB\ model since the structures become so strongly 
nonlinear that shell crossings occur before $S/N = 1$ is reached.

   The ``kSZ wedge'' has an interesting interpretation: the kSZ effects due 
to the near and far sides of structures entirely at $z < z_{\rm re}$, for 
reionization at $z_{\rm re}$, approximately cancel.  For structures 
centred near $z_{\rm re}$, however, very little kSZ is generated at the 
far side, so there is no cancellation, leading to strong constraints 
along the wedge.

   Figure~\ref{psilimplot} shows at a glance how large the amplitude can be 
for structures that are hiding in our observable volume, as a function of 
size and distance, and in particular outside the region of parameter space 
directly sampled with galaxy surveys.  For example, a $1$~Gpc-radius 
structure halfway to our LSS with as much as $15$ times the expected \LCDM\ 
amplitude would go unnoticed \cite{noteCopern}.

   Repeating our calculations for an {\em over}density with profile 
(\ref{profile}), we find very small differences due to the change in 
sign between the RS and other anisotropies.  Underdensities with the 
non-compensated profile from \cite{mzs11} result in somewhat weaker 
constraints in some areas and stronger constraints in others, with the 
broad picture unchanged.


\section{Statistical inhomogeneity}

   Next we address the limits 
that can be placed on statistical inhomogeneity from CMB secondaries as well 
as galaxy lensing data.  Our approach is to constrain separately the 
primordial spectrum using primary anisotropies at $z > 100$, 
$\P_\R^{\rm LS}(k)$, and the primordial spectrum using secondaries and 
galaxy lensing at $z \le 100$, $\P_\R^{\rm in}(k)$.  As \fig\ref{krplot} 
shows, $z = 100$ separates reasonably well the primary and secondary 
sources.  The last scattering (LS)  spectrum is parameterized as usual by 
$\P_\R^{\rm LS}(k) = A_{\rm S}(k/k_0)^{n_{\rm S} - 1}$ (with pivot scale 
$k_0 = 0.05\,{\rm Mpc}^{-1}$).  We parameterize $\P_\R^{\rm in}(k)$ by a 
cubic spline, choosing 7 knots a decade apart from $k = 10^{-5}$ to 
$10\,{\rm Mpc}^{-1}$, so all observable scales are well within this range.  
Our approach is sensitive to statistical inhomogeneity, with different 
primordial spectra inside versus at the LSS, and serves as a consistency test 
for \LCDM.

   The data we use are: the CMB temperature power spectrum 
measured by \Planck~\cite{planck13params}, with polarization and 
temperature-polarization ($TE$) cross-correlation power spectra for 
$\ell\leq 32$ from the WMAP9 release~\cite{2013ApJS..208...19H}; the 
\Planck\ lensing potential power spectrum~\cite{planck13lensing}; and the 
2D cosmic shear signal measured by the Canada-France-Hawaii Telescope 
Lensing Survey~\cite{2013MNRAS.430.2200K}.  We restrict the shear signal to 
linear scales by imposing an angular cut of $\theta_c=17'$ and $53'$ 
for the $\xi^{+}$ and $\xi^{-}$ correlation functions, respectively 
(see~\cite{2013MNRAS.430.2200K} for details). The Markov-Chain-Monte-Carlo 
fitting methodology requires some 
modification with our approach: the transfer functions and power spectrum 
are computed twice, once restricting the source functions to  $z > 100$ 
and once to  $z \le 100$, and each component of the total likelihood is 
then given the appropriate spectrum.  This approach works insofar as one 
can isolate the observables to each redshift range: this is a good 
approximation, apart from the {\em lensed} primary temperature power 
spectrum at high $\ell$.  We take  $\P_\R^{\rm LS}(k)$ 
to smooth the primary CMB spectrum, so our results are conservative by 
neglecting any modification to this.  Note that ISW, reconstructed 
lensing potential, and galaxy lensing {\em are} modified by the interior 
power spectrum $\P_\R^{\rm in}(k)$.  Also, note that we automatically 
include the (large-scale) kSZ effect via the Boltzmann evolution.

\begin{figure}[t]
\includegraphics[width=\columnwidth]{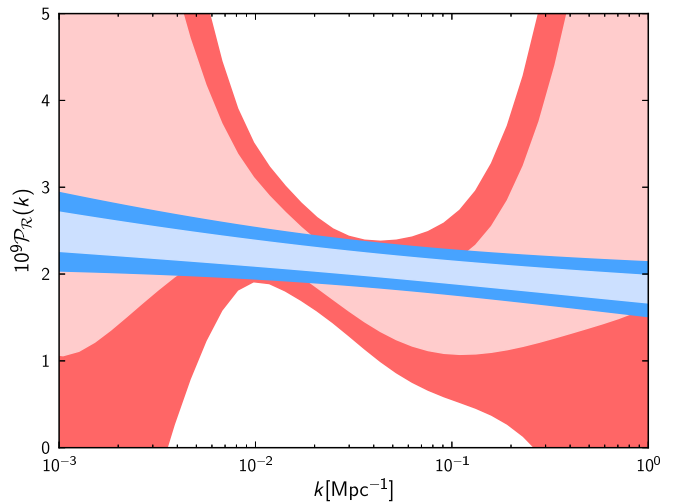}
\caption{1 and 2$\sigma$ confidence intervals of the primordial spectrum 
$\P_\R^{\rm LS}(k)$ for $z > 100$ (blue) and $\P_\R^{\rm in}(k)$ for 
$z \le 100$~(red).}
\label{PSinhomplot}
\end{figure}

   The results of our analysis are illustrated in Fig.~\ref{PSinhomplot}, 
where we show the 1 and 2$\sigma$ confidence intervals of the LS and 
interior primordial power spectra.  Unsurprisingly, the LS spectrum is 
more tightly constrained, although the errors would increase 
somewhat if the power law assumption was relaxed.  The interior spectrum 
is consistent with the LS spectrum, and is constrained quite well 
in the range $k \simeq 0.008$ to $0.2\,{\rm Mpc}^{-1}$, precisely the 
scales probed by CMB and galaxy 
lensing.  Larger scales are more weakly constrained by CMB lensing, ISW, 
and $TE$ polarization.  $\P_\R^{\rm in}(k)$ is pulled down relative to 
$\P_\R^{\rm LS}(k)$ by the galaxy lensing data on small scales and pushed 
up by CMB lensing on larger scales.  
One consequence of the additional freedom in 
$\P_\R^{\rm in}(k)$ is an increase in the errors on other cosmological 
parameters.  With only $TE$ data, e.g., the optical depth to 
reionization, $\tau$, would be completely degenerate with $A_{\rm S}$.  
However, this degeneracy is broken with the inclusion of temperature data, 
and we find $\tau = 0.06 \pm 0.03$.


\section{Conclusions}

   The results in this paper represent our first 
quantitative look at what might be ``hiding'' in most of our observable 
volume.  
Figure~\ref{psilimplot} quantifies how large departures from homogeneity, 
in the form of spherical structures, are allowed by current data.  
In particular, CMB lensing (and to a lesser 
extent ISW and kSZ) restrict structures over most of our observable 
volume to $\lesssim$~20 times the \LCDM\ amplitude.  We also find that, 
while the ISW effect is a good probe of very large scales at late times, 
CMB lensing is unmatched in its ability to inform us about most of the 
visible Universe and a wide range of scales.

   Similarly, \fig\ref{PSinhomplot} illustrates our current 
ability to constrain statistical inhomogeneity in the fluctuation spectrum 
which affects the secondary but not the primary CMB.  Equivalently, 
this represents the first power spectrum reconstruction based on CMB 
secondaries alone.  \LCDM\ has passed an important new consistency test.

   There are many possibilities for followup work.  For localized 
structures, the assumption of spherical symmetry could 
be relaxed using linear perturbation theory, at least outside the region of 
parameter space where nonlinear structures are allowed.  For statistical 
inhomogeneity, there is clearly much freedom in how to parameterize 
the interior power spectrum, since not only could the radial 
distribution be modified, but non-trivial angular dependence should be 
considered as well (motivated, perhaps, by suggestions of a dipolar 
CMB power asymmetry \cite{planck13isostat} that would actually extend 
{\em to} our LSS).

   In the near future, we can expect improved CMB lensing measurements from 
the South Pole Telescope, and also from \Planck\ with the availability 
of the full mission data, including polarization.  Farther ahead, we 
can hope that information about fluctuations over very large distances 
can be retrieved from remote CMB quadrupole measurements \cite{bunn06} or 
from other potential probes \cite{maartens11,clarkson12}.  
Ultimately, we can anticipate that surveys will again take the lead, 
as 21-cm measurements may map most of our observable volume \cite{pl12} and 
provide a nearly complete picture of our patch of the Universe.

%
%



%
%



\section*{Acknowledgments}

   We thank D. Hanson for assistance with the \Planck\ lensing data and 
D. Scott for comments on the draft.  This research was supported by the 
Canadian Space Agency and the STFC. We acknowledge the use of the 
{\tt CAMB}~\cite{2000ApJ...538..473L} and 
{\tt COSMOMC}~\cite{2002PhRvD..66j3511L} codes.

\bibliography{homog}

\begin{thebibliography}{70}
\bibitem[{not({\natexlab{a}})}]{noteisotr}
\bibinfo{note}{Note that the connection between isotropy of the CMB and
  isotropy of the spacetime is not trivial (see, e.g., \cite{maartens11}).}
\bibitem[{not({\natexlab{b}})}]{noteinfl}
\bibinfo{note}{Another reason is the theoretical bias that we expect inflation
  to homogenize and isotropize the Universe, leaving only the fluctuations
  generated by inflation.}
\bibitem[{not({\natexlab{e}})}]{noteell}
\bibinfo{note}{For power spectrum $\ell^nC_\ell$, the choice of $n$ is to an
  extent arbitrary. Our choices are conventional, but different values will
  shift the peak of the distributions in $\ell$ somewhat.}
\bibitem[{not({\natexlab{f}})}]{noteradn}
\bibinfo{note}{Our approach is accurate during matter or $\Lam$ domination, but
  radiation will affect the accuracy at the largest distances. E.g., for $r >
  0.97r_{\rm LS}$, we have $\rho_{\rm rad}/\rho_m > 0.1$ in the standard model.
  An early ISW effect is expected to modify our results slightly at the largest
  $r$ values, where the constraints are strongest due to the SW effect.}
\bibitem[{not({\natexlab{i}})}]{noteampl}
\bibinfo{note}{Note that this expected amplitude is only defined up to a
  constant factor of order unity, since it must depend on the shape of the
  structure's profile.}
\end{thebibliography}

\end{document}